\documentclass[12pt]{article}
\usepackage{scicite}
\usepackage{times}
\usepackage{graphicx}

\newenvironment{sciabstract}{%
\begin{quote} \bf}
{\end{quote}}

\newcounter{lastnote}
\newenvironment{scilastnote}{%
\setcounter{lastnote}{\value{enumiv}}%
\addtocounter{lastnote}{+1}%
\begin{list}%
{\arabic{lastnote}.}
{\setlength{\leftmargin}{.22in}}
{\setlength{\labelsep}{.5em}}}
{\end{list}}

\title{The Cosmic Production of Helium}
\author{Raul Jimenez,$^{1\ast}$ Chris Flynn,$^{2,3}$ James MacDonald$^4$ and Brad K. Gibson $^3$\\
\\
\normalsize{$^{1}$Department of Physics and Astronomy, University of Pennsylvania, PA 19104-6396, USA}\\ 
\normalsize{$^{2}$Tuorla Observatory, Piikki\"o, FIN-21500, Finland}\\ 
\normalsize{$^{3}$Centre for Astrophysics \& Supercomputing, Swinburne
University, Victoria, Australia}\\ 
\normalsize{$^{4}$Department of Physics and
Astronomy, University of Delaware, Newark DE 19716, USA}\\
\normalsize{$^\ast$To whom correspondence should be addressed; E-mail: raulj@physics.upenn.edu}
}

\date{}

\begin{document}

\maketitle 

\begin{sciabstract}
  We estimate the cosmic production rate of helium relative to metals
  ($\Delta Y/\Delta Z$) using K dwarf stars in the Hipparcos catalog
  with accurate spectroscopic metallicities. 
  The best fitting value is $\Delta Y/\Delta Z=2.1 \pm 0.4$ at the
  68\% confidence level.
  Our derived value agrees with determinations from H{\sc II} regions
  and with theoretical predictions from stellar yields with standard
  assumptions for the initial mass function. The amount of helium in
  stars determines how long they live and therefore how fast they will
  enrich the insterstellar medium with fresh material.
\end{sciabstract}

The amount of helium divided by the amount of heavier elements
produced in stars ($\Delta Y/\Delta Z$), is of great interest to
astrophysics and cosmology. The ratio governs the stellar clock and
thus how long stars will live. Therefore, age determinations of both
resolved and integrated stellar populations rely on knowing how helium
concentration changes as a function of metallicity. Because ages of
galaxies can help to determine the nature of dark energy \cite{JL02}, an
accurate determination of $\Delta Y/\Delta Z$ is needed.  It also has
an impact on the determination of primordial helium abundance by use
of extra-galactic H{\sc II} regions \cite{PT76}. In addition to the
above cosmological interest, it is also a test of theoretical
predictions of stellar yields, because given an initial mass function
for stars, $\Delta Y/\Delta Z$ is a predicted quantity of stellar
evolution.

Helium lines can be observed only in stars with an effective
temperature greater than $20,000 K$; it is not possible to measure
helium abundances directly in cooler stars which live more than $10^8$
yr. There are two methods to determine observationally the value of
$\Delta Y/\Delta Z$ (where $Y$ and $Z$ are the fractional abundances
of helium and all metals hevier than helium, respectively). One is
through the use of H{\sc II} extra-galactic regions to measure
helium's primordial abundance, which then is compared to that of the
Sun to deduce $\Delta Y/\Delta Z$. This is done by either
extrapolation of correlations of $Y-(O/H)$ with $Y-(N/H)$ to
$O/H=N/H=0$ (e.g. \cite{ITL97,PPL02,GSV02} and references therein) or
by measuring $Y$ in ultra-low metallicity blue compact dwarf galaxies
and assuming that no chemical evolution has taken place (e.g.
\cite{ICG01}).  Recent values for $\Delta Y/\Delta Z$ range from 2.5
\cite{TI02} to 3.5 \cite{PPR00}, which agrees with
stellar yield computations assuming a standard Salpeter, Scalo or
Kroupa initial mass function (IMF) \cite{Maeder92,Tsujimoto+97}.

The other (indirect) method exploits the fact that the temperature of
main sequence dwarf stars is insensitive to stellar age. The strongest
dependences are on $Y, Z$ (which can be determined from spectroscopy)
and on the value of the mixing length parameter. The mixing length
parameter can be calibrated in stellar models using the Sun and the
very thin and narrow red giant branches of globular clusters indicate
that other stars have the same mixing length as the Sun. Further, the
fact that all globular clusters require the same value for the mixing
length parameter supports the supposition that the mixing length
parameter does not depend on $Z$ (e.g. \cite{Jimenez+96}). Although
rotation may affect the temperature of a main sequence star, its
effect is small \cite{Fernandes+96}. Therefore, once the metallicity
has been measured, the color and luminosity of the main sequence
depends only on the value of $Y$. Because the isochrone for the Sun is
a fixed point (the $Y$ value is adjusted to match the present age and
radius of the Sun), the effect of varying $\Delta Y/\Delta Z$
anti-correlates with the broadening of the main sequence stellar
values. Here we apply a related method to that employed by
\cite{PagelPortinari98} and newly available spectroscopic
metallicities of stars in the Hipparcos sample to put tighter
constraints on the value of $\Delta Y/\Delta Z$.

This method of determining $\Delta Y/\Delta Z$ has been applied to
stars in the solar neighborhood using ground-based parallaxes.
\cite{Perrin+77} found a value $\Delta Y/\Delta Z = 5 \pm 3$,
\cite{Fernandes+96} found $\Delta Y/\Delta Z > 2$. Recently,
\cite{PagelPortinari98} used the Hipparcos parallaxes, infrared flux
temperatures and accurate, spectroscopically determined metallicities
for sub-solar metallicity stars. They derived a value $\Delta Y/\Delta
Z = 3 \pm 2$.

The sample consists of K dwarfs for which we have accurate Hipparcos
parallaxes, accurate spectroscopically determined metallicities, and
broad band in filters $V$ and $B$.  Our primary sources of
metallicities are G and K dwarfs in
\cite{Flynn_Morell_97,ThorenFeltzing00,FeltzingGonzalez01,Chaboyer+98,TomkinLambert99}.
No binaries or suspected binaries were used. The complete list of
sample stars is shown in Table~1.

We only use stars in the absolute magnitude range $5.5 < M_V < 7.5$.
The limit at $M_V = 5.5$ ensures the stars have sufficiently low mass
that no significant stellar evolutionary effects have taken place
since they arrived on the main sequence; stars more luminous than this
limit have begun to evolve off the main sequence, meaning that their
luminosities are no longer primarily dependent on 
$Z$ and $Y$. Stars fainter than $M_V = 7.5$ are
excluded because metallicity determinations for these cooler, lower
mass stars are still difficult to perform because of their complex
spectra. The $B-V$ colors mainly come from \cite{Bessell90}, our own
photometry \cite{KFJ02} and the Hipparcos catalog.

The absolute magnitude of a star relative to the luminosity of the
solar metallicity isochrone at the same $B-V$ color is defined as
$\Delta M_V$ (figure~1). This quantity is tightly correlated with
metallicity, as first shown in \cite{KFJ02}, as seen in figure~2. 

For a fixed $Z$, an increase in $Y$ translates into an
increase of luminosity and effective temperature due to the increase
in molecular weight. On the other hand, an increase in $Z$
produces a dimming and cooling of the star due to the presence of more
metals. Although this behavior can be understood by using
quasi-homology relations which are derived for fully radiative cores
(e.g. \cite{Fernandes+96}), it is not useful because dwarfs
along the main sequence have rather deep convective envelopes.
We have used the stellar evolution code JMSTAR \cite{JM96,MM01,LM03}
to compute isochrones for $5.5 < M_V < 7.5$ and $0.4 < B-V < 1.4$, or
equivalently effective temperatures, $6000 < T_{\rm eff} < 4500 $.  We
use of opacities from \cite{Alexander_Ferguson_94} for high $Z$ stars,
because otherwise these stars would be too blue
\cite{Castellani+01}. We used the code to compute stellar tracks on
the main sequence for stars with masses $0.6 < M/M_{\odot} < 1.2$. We
have computed isochrones from the above tracks and then converted them
from the theoretical ($L-T_{\rm eff}$) to the observational plane
($M_V-(B-V)$) using the most recent Kurucz atmospheres
models\footnote{Available at {\tt
http://cfaku5.harvard.edu/grids.html}. Recently, \cite{BCP98} have
studied the reliability of these models, concluding they do an
excellent job at predicting the correct color$-T_{\rm eff}$ relation
when compared with other methods (e.g the infrared flux method).}.

We use the fiducial solar isochrone (Fig.~1) as our reference point to
measure $\Delta M_V$ for the model stars at a fixed $B-V$ color.  The
locus of the stars in the $\Delta M_V$ versus log$Z$ plane shows a
correlation between luminosity and metallicity that is consistent with
the theoretical stellar evolutionary predictions for isochrones with
$\Delta Y/\Delta Z=1, 2$ and $4$. While the luminosity of the stars is
primarily sensitive to metallicity, we also confirm our expectation
that the effect of $Y$ is only seen when the metallicity is
sufficiently high ($Z/Z_{\odot} > 0.3)$.  Note that we have chosen the
primordial abundance to be $Y_p = 0.236$ \cite{PPL02}, so that at
$Z/Z_{\odot} = 0.3$ the corresponding values of $Y$ are 0.24, 0.25 and
0.26 for $\Delta Y/\Delta Z = 1$, 2 and 4, respectively: i.e. these
are sufficiently different in $Y$ to start to have an effect on the
luminosity of the star. We have chosen to fix $Y_p$ and then compute
the corresponding $Y$ by using $Y=Y_p+\frac{\Delta Y}{\Delta Z}Z$,
instead of what \cite{PagelPortinari98} do, which is to compute $Y$
using the Sun as the anchor point. The results of
\cite{PagelPortinari98} lead to primordial helium abundance of 0.26
and 0.20 for $\Delta Y/\Delta Z = 1$ and 4, respectively which are
inconsistent with values inferred from H{\sc II} regions
\cite{ITL97,ICG01,PPL02,GSV02}. Even if the primordial value has been
enhanced due to the presence of a primordial zero metallicity
population with very massive stars (up to 1000 M$_{\odot}$), the
effect will be very small \cite{Marigo+02}, only about 0.01, which
does not affect our conclusions.

The isochrones with $\Delta Y/\Delta Z = 4$ are not favored by our
analysis. To obtain the statistical significance of the best fitting
model, we have performed a maximum likelihood fit of the data to the
models. Because the errors are Gaussian distributed and uncorrelated,
this translates into simple $\chi^2$ statistics.  We found the best
$\chi^2$ by minimizing the distance of the observational points to the
models (Fig.~3). The reduced $\chi^2$ for $\Delta Y/ \Delta Z = 2$ is
1.2, which indicates that these models are reasonable fit. The
probability of this value of $\chi^2$ being the correct model, using
the incomplete gamma function of the number of degrees of freedom and
the value of $\chi^2$, is about 20\%. We computed 4 models for $\Delta
Y / \Delta Z = 1$, 2, 3 and 4 and fitted a Gaussian to the likelihood,
which is a parabola in $\chi^2$, to these points (Fig.~3). Both values
1 and 4 are excluded at greater than the $2\sigma$ level.  The best
fit corresponds to $\Delta Y/\Delta Z = 2.1 \pm 0.4$. Which represents
a much narrower range than previous studies. The low metallicity part
($\log_{10} Z < -2.4$) of the sample can be used to determine the
value of primordial helium $Y_p$, because this part is not sensitive to
$\Delta Y/ \Delta Z$ (Fig.~2). A maximum likelihood fit produces
$Y_p=0.24 \pm 0.02$, which is consistent with our chosen value and
those derived from HII regions but with a larger error
\cite{ITL97,ICG01,PPL02,GSV02} .

The dominant systematic uncertainty plaguing our analysis can be
traced to the inherent assumption of solar-scaled abundances in both
the data and theoretical stellar atmospheres. Only a small fraction of
the super-solar metallicity K-dwarfs in the solar neighborhood are
enhanced in $\alpha$-elements with respect to iron \cite{Taylor70}.
The $\alpha$-enhanced isochrones of \cite{Salasnich+00} demonstrate
that for every 0.2 dex increase in [$\alpha$/Fe] above the
scaled-solar ratio, $\Delta$M$_{\rm V}$ increases by $\sim$0.07\,mag.
Our sample does not have such a high enhancement; multi-species
abundances \cite{ThorenFeltzing00,FeltzingGonzalez01} show a maximum
value of 0.05 with an average value of zero, consistent with no
enhancement. As such, we assign a conservative systematic error budget
on $\Delta$M$_{\rm V}$ of $^{+0.0}_{-0.02}$. On the contrary, low
metallicity stars have [$\alpha$/Fe]$\sim$0.4. For metal-poor stars
the effect of alpha enhancement can be accurately approximated by
scaling to a higher metallicity the solar-scaled isochrones
\cite{PagelPortinari98}. The effect this has on an isochrone is to
make it cooler but also brighter (since we are measuring $\Delta M_v$
at a fixed color). These two compensate each other and the net effect
is that the points slide along the solar-scaled isochrones of figure 2.

We derive $\Delta Y/\Delta Z = 2.1 \pm 0.4$ at the 68\% confidence
level. Our $\Delta Y/\Delta Z$ agrees with modern determinations from
H{\sc II} regions ($Y_p=0.236$)\cite{ICG01,PPL02,GSV02} and the helium
abundance needed to fit the Sun with its current age and radius
($Y_{\odot}=0.275$, $Z_{\odot}=0.017$), which gives $\Delta Y/\Delta
Z=2.3$. It also agrees with theoretical predictions from stellar
yields with standard assumption for the initial mass function
\cite{Maeder92,Tsujimoto+97}, which translates into a upper mass
cut-off of 100 M$_{\odot}$ for a Salpeter IMF. But, is not consistent
with $\Delta Y/\Delta Z$ derived for the Hyades \cite{LFL01}, which is
$<1$. If the Hyades value is confirmed, it will indicate a different
evolutionary path in the chemical enrichment in the Milky Way.


\begin{scilastnote}
\item RJ is supported by NSF grant AST-0206031. BKG \& CF acknowledge
  the support of the Australian Research Council through its Large
  Research Grant (A010517), Discovery Project (DP0343508), and Linkage
  International Award (LX0346832) schemes. This research was
  supported by the Academy of Finland through its funding of the
  ANTARES program for space research. We have made extensive use of
  the Simbad stellar data base, at the Centre de Donn\'ees
  astronomiques de Strasbourg, for which we are very grateful.
\end{scilastnote}

\clearpage

\begin{figure}
\includegraphics[width=15cm]{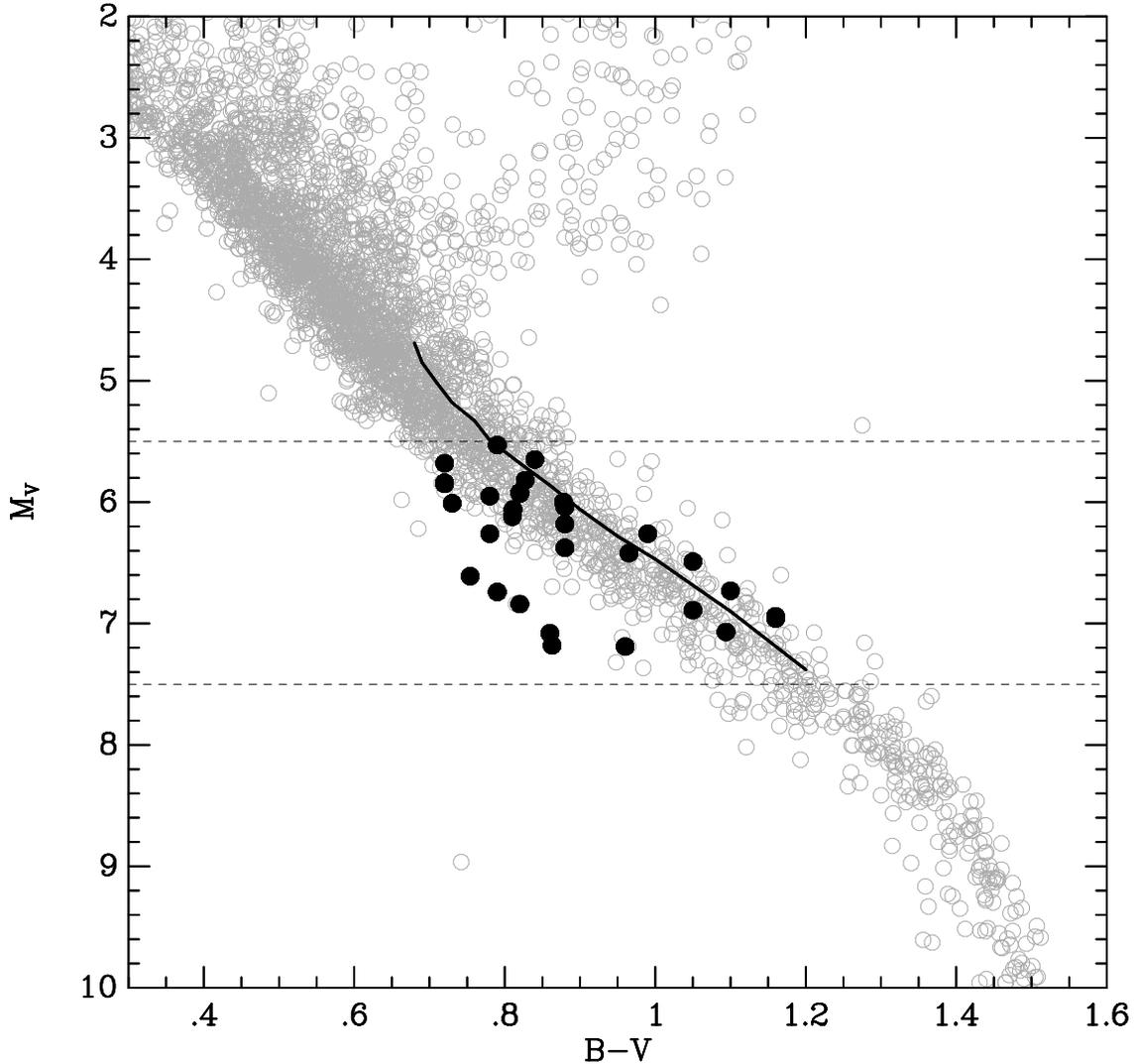}
\caption{Our sample of K dwarf stars (solid symbols) plotted over the Hipparcos
  color-magnitude diagram (open symbols).  The reference solar
  isochrone is shown by the solid line. The difference
  between a star's luminosity $M_V$ and the luminosity of the
  isochrone at the same color is the quantity $\Delta M_V$, which
  correlates tightly with the stellar metallicity in figure~2. Note
  that in plotting Hipparcos stars (open symbols), only non-multiple
  stars with parallax errors smaller than 5\% and color errors less
  than 0.02 magnitudes have been used. There are a small number of
  stars a few tenths of a magnitude above the main
  sequence which are likely to be unrecognized binaries.}
\label{cmd}
\end{figure}

\clearpage

\begin{figure}
\includegraphics[width=16cm]{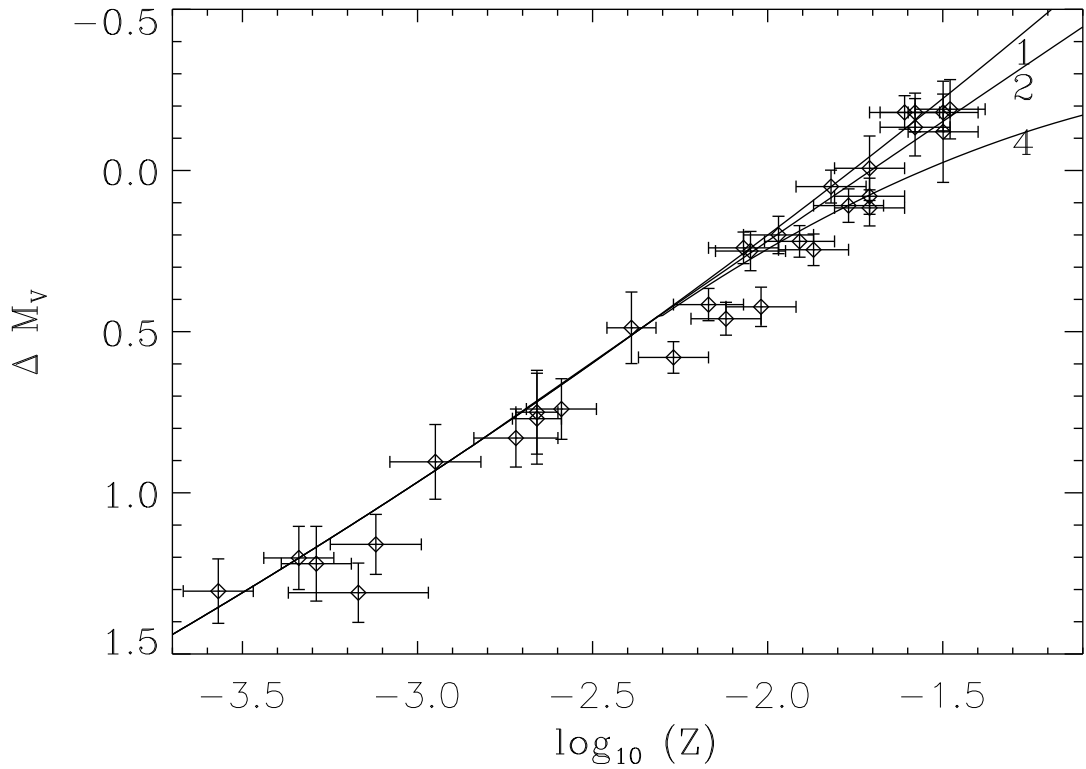}
\caption{The solid lines correspond to $\Delta M_V$, measured from the
  fiducial solar isochrone, from theoretical models for K dwarfs for
  different values of $\Delta Y/ \Delta Z$ (1, 2 and 4 from top to
  bottom) as a function of metallicity. The observational points are
  for K dwarfs with Hipparcos parallaxes and with spectroscopically
  determined metallicities. The errors in $\Delta M_V$ are driven by
  the color and parallax errors in about equal proportion.  The
  spectroscopically determined metallicities have errors of typically
  0.1 dex.}
\label{zcorr}
\end{figure}

\clearpage

\begin{figure}
\includegraphics[width=16cm]{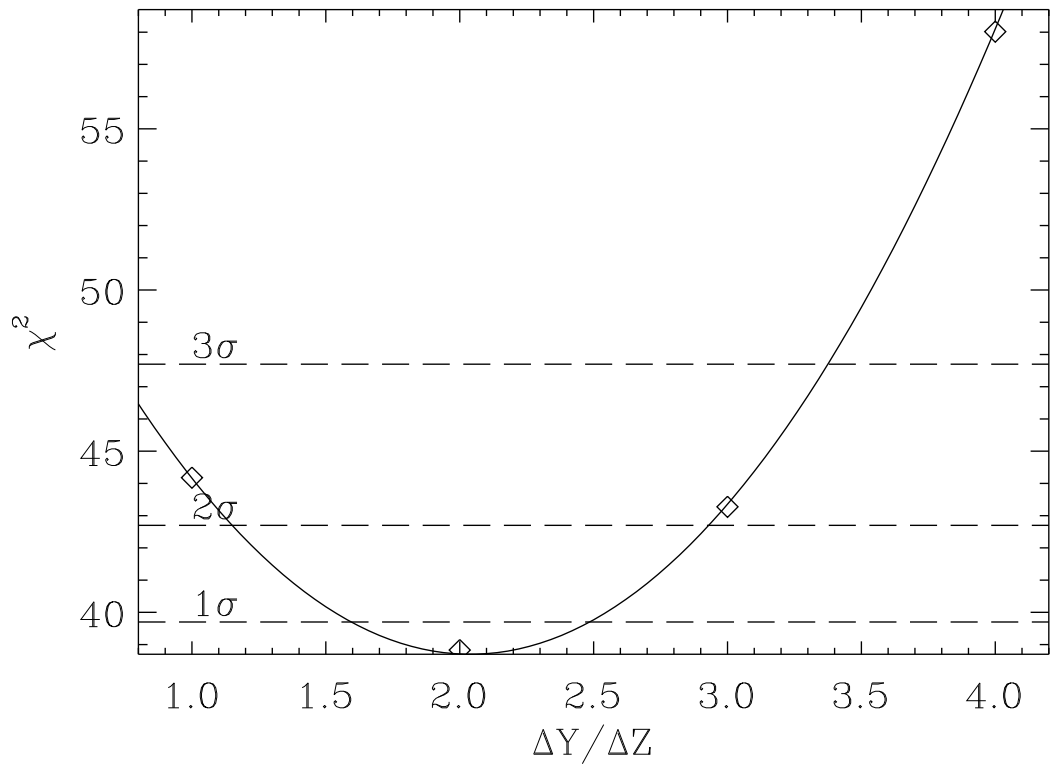}
\caption{$\chi^2$ for four models (diamonds) and the data of figure~2. 
  The solid line is a parabola fitted to the models. The dashed lines
  represent several confidence levels. Note that values as low as 1
  and as high as 3 are excluded at more than 95\% confidence.}
\label{maxlhood}
\end{figure}

\clearpage

\begin{table}
\begin{center}
\caption{The sample of Hipparcos K dwarfs with spectroscopic metallicities. The
final column ``S'' indicates the sources of the metallicities. The codes are:
1: Tomkin and Lambert (1999). 2: Thoren and Feltzing (2000), Feltzing and
Gonzalez (2001). 3: Chaboyer et al (1998). 4: Flynn and Morell (1997). 5:
Cayrel catalog.}
\vspace*{0.2cm}
\begin{tabular}{lrrrrrrrr}
\hline
   Name    &       $V$ & $B-V$     & $M_V$     & [Fe/H]      & $\Delta M_V$  &  S \\  
\hline
  HD 4628 &      5.74 &      0.88 &      6.38 & $    -0.40$ & $      0.42 $ & 4\\
  HD 10700 &      3.53 &      0.72 &      5.68 & $    -0.50$ & $      0.58 $ & 4\\
  HD 13445 &      6.12 &      0.82 &      5.93 & $    -0.28$ & $      0.25 $ & 4\\
  HD 21197 &      7.86 &      1.16 &      6.96 & $     0.27$ & $     -0.23 $ & 2\\
  HD 22049 &      3.72 &      0.88 &      6.18 & $    -0.14$ & $      0.22 $ & 1\\
  HD 25329 &      8.50 &      0.86 &      7.18 & $    -1.80$ & $      1.30 $ & 5\\
  HD 26965 &      4.41 &      0.82 &      5.92 & $    -0.30$ & $      0.24 $ & 4\\  
  HD 30501 &      7.58 &      0.88 &      6.04 & $     0.06$ & $      0.08 $ & 2\\
  HD 31392 &      7.60 &      0.79 &      5.53 & $     0.06$ & $     -0.01 $ & 2\\
  HD 32147 &      6.22 &      1.05 &      6.49 & $     0.29$ & $     -0.19 $ & 4\\
  HD 61606 &      7.18 &      0.96 &      6.42 & $     0.06$ & $      0.08 $ & 2\\
  HD 64606 &      7.43 &      0.73 &      6.01 & $    -0.95$ & $      0.83 $ & 1\\
  HD 65583 &      6.97 &      0.72 &      5.84 & $    -0.82$ & $      0.74 $ & 1\\
  HD 72673 &      6.38 &      0.78 &      5.95 & $    -0.35$ & $      0.46 $ & 4\\
  HD 87007 &      8.82 &      0.84 &      5.65 & $     0.27$ & $     -0.12 $ & 2\\
  HD 100623 &      5.96 &      0.81 &      6.06 & $    -0.25$ & $      0.42 $ & 4\\
  HD 103095 &      6.45 &      0.75 &      6.61 & $    -1.40$ & $      1.31 $ & 3,4\\
  HD 103932 &      6.99 &      1.16 &      6.94 & $     0.16$ & $     -0.25 $ & 2\\
  HD 108564 &      9.45 &      0.96 &      7.19 & $    -1.18$ & $      0.87 $ & 1\\
  HD 134439 &      9.10 &      0.79 &      6.74 & $    -1.57$ & $      1.20 $ & 4\\
  HD 134440 &      9.45 &      0.86 &      7.08 & $    -1.52$ & $      1.22 $ & 4\\
  HD 136834 &      8.28 &      0.99 &      6.26 & $     0.19$ & $     -0.17 $ & 2\\
  HD 145417 &      7.53 &      0.82 &      6.84 & $    -1.35$ & $      1.16 $ & 5\\
  HD 149661 &      5.76 &      0.83 &      5.82 & $     0.00$ & $      0.11 $ & 4\\
  HD 192031 &      8.67 &      0.72 &      5.85 & $    -0.89$ & $      0.75 $ & 1\\
  HD 192310 &      5.73 &      0.88 &      6.00 & $    -0.05$ & $      0.05 $ & 4\\
  HD 209100 &      4.69 &      1.05 &      6.89 & $    -0.10$ & $      0.21 $ & 4\\
  HD 213042 &      7.67 &      1.10 &      6.73 & $     0.19$ & $     -0.17 $ & 2\\
  HD 216803 &      6.48 &      1.09 &      7.07 & $    -0.20$ & $      0.20 $ & 4\\
BD +41 3306 &      8.86 &      0.81 &      6.12 & $    -0.62$ & $      0.49 $ & 1\\
BD +24 4460 &      9.51 &      0.78 &      6.26 & $    -0.89$ & $      0.77 $ & 1\\
\hline
\end{tabular}
\label{tab1}
\end{center}
\end{table} 

\end{document}